\documentclass[nofootinbib,superscriptaddress,twocolumn,showpacs,preprintnumbers,pre,aps]{revtex4-1}
\usepackage{graphicx}
\usepackage{bm}
\usepackage{amsmath}
\usepackage{amssymb}
\usepackage{color}
\newcommand{\bmrm}[1]{\bm{\mathrm{#1}}}

\begin{document}

\title{Thermal and active fluctuations of a compressible bilayer vesicle}

\author{T. V. Sachin Krishnan}
\affiliation{Department of Physics, Indian Institute of Technology Madras,
Chennai, 600036, India}
\affiliation{Department of Chemistry, Graduate School of Science and Engineering,
Tokyo Metropolitan University, Tokyo 192-0397, Japan}

\author{Kento Yasuda}
\affiliation{Department of Chemistry, Graduate School of Science and Engineering,
Tokyo Metropolitan University, Tokyo 192-0397, Japan}

\author{Ryuichi Okamoto}
\affiliation{Research Institute for Interdisciplinary Science, Okayama University,
Okayama 700-8530, Japan}

\author{Shigeyuki Komura}\email{komura@tmu.ac.jp}
\affiliation{Department of Chemistry, Graduate School of Science and Engineering,
Tokyo Metropolitan University, Tokyo 192-0397, Japan}

\date{\today}
%\date{February 9, 2018}

\begin{abstract}
We discuss thermal and active fluctuations of a compressible bilayer vesicle by using the 
results of hydrodynamic theory for vesicles.
Coupled Langevin equations for the membrane deformation and the density fields are employed 
to calculate the power spectral density matrix of membrane fluctuations.
Thermal contribution is obtained by means of the fluctuation dissipation theorem, whereas active 
contribution is calculated from exponentially decaying time correlation functions of active 
random forces.
We obtain the total power spectral density as a sum of thermal and active contributions.
An apparent response function is further calculated in order to compare with the recent 
microrheology experiment on red blood cells. 
An enhanced response is predicted in the low-frequency regime for non-thermal active fluctuations.
\end{abstract}

\maketitle

%%%%%%%%%%%%
\section{Introduction}
%%%%%%%%%%%%

Vesicles idealized as closed two-dimensional (2D) elastic sheets serve as a simple model for 
complex biological cells such as the red blood cells (RBCs).
Studies of vesicle fluctuations have offered significant insights into the physical 
properties of RBCs.
In thermal equilibrium, local deformations of lipid membranes obey well-known statistics 
prescribed by several physical parameters~\cite{SafranBook}.
Out-of-plane membrane fluctuations were measured by flicker analysis~\cite{Rodriguez-Garcia09,Boss12} 
to obtain both static and dynamic properties such as the membrane bending 
modulus~\cite{Faucon89} or the wavelength-dependent relaxation rates~\cite{Milner87}.

Since the deformation energy of a lipid membrane is comparable to thermal energy, earlier 
theoretical works assumed that vesicle fluctuations are driven by thermal forces~\cite{Brochard75}.
However, recent experiments have revealed the existence of active non-thermal fluctuations
in addition to thermal fluctuations~\cite{Betz09, Park10}. 
Using both active and passive microrheology techniques, Turlier \textit{et al.}\ measured the power 
spectral density (PSD) as well as the response function of a single RBC~\cite{Turlier16}.
Moreover, they demonstrated a violation of the fluctuation dissipation theorem (FDT) due 
to non-thermal active forces. 
This breakdown of FDT seems to be caused by metabolic processes taking place inside the 
RBC such as the pumping action of ion channels or the growth of cytoskeletal networks~\cite{Turlier16}.

Recent experiments have further clarified the functional significance of ATP-driven non-thermal 
fluctuations in cells~\cite{Biswas17}.
It was shown that the activity of motor protein F$_1$F$_0$-ATPase and actomyosin 
cytoskeleton affect the local organization as well as the elastic properties of the lipid 
membrane~\cite{Almendro-Vedia17}.
Moreover, membrane fluctuations act as a generic regulatory mechanism in cell 
adhesion by controlling the interaction between cadherin proteins~\cite{Fenz17}.

So far, several theoretical models have been suggested to describe the statistics of membrane 
fluctuations in the presence of active forces.
Prost and Bruinsma proposed a model for membrane ion pumps and showed that active fluctuations
dominate at large wavelengths~\cite{Prost96}.
Later, Ramaswamy \textit{et al.}\  considered a model that accounts for the coupling between 
the ion pump density and the membrane curvature, which can destabilize the membrane 
shape~\cite{Ramaswamy00}.
For membranes containing ion pumps, their time-dependent fluctuation or the mean square 
displacement (MSD) was later obtained by Lacoste and some of the present 
authors~\cite{Lacoste05,Komura15}.
Apart from ion pumps in membranes, cytoskeletal proteins also act as sources of active forces 
which affect the membrane fluctuation spectrum in a significant 
manner~\cite{Gov04,Gov05, Gov07,Yasuda16}.

One important aspect that has been often neglected in the dynamics of a bilayer membrane is 
the role of inter-monolayer friction and hence the membrane lateral compressibility. 
As a consequence of the bilayer structure, lipid monolayers are inevitably compressible in 
order to allow out-of-plane membrane deformations.
It is known that the compression mode due to inter-monolayer friction drastically 
affects the bilayer dynamics~\cite{Seifert93,Miao02,Fournier15,Okamoto16}.
Recently, the present authors have investigated the relaxation dynamics of a compressible 
bilayer vesicle with an asymmetry in the viscosity of the inner and outer fluid 
medium~\cite{SachinKrishnan16}. 
Using the framework of the Onsager's variational principle~\cite{DoiBook}, we showed that 
different relaxation modes are coupled to each other due to the bilayer structure of a spherical vesicle. 
We have also discussed the dynamics of a bilayer membrane coupled to a 2D
cytoskeleton, for which the slowest relaxation is governed by the intermonolayer 
friction~\cite{Okamoto17}. 
It was predicted that forces applied at the scale of cytoskeleton for a sufficiently long 
time can cooperatively excite large-scale modes.

In this paper, we discuss both thermal and active fluctuations of a compressible bilayer vesicle.
Using the results of hydrodynamic theory that explicitly takes into account the 
intermonolayer friction~\cite{SachinKrishnan16}, we consider the standard Langevin equation 
for the membrane deformation and the lipid density 
fields neglecting hydrodynamic memory whose effects will be separately discussed in the final section.
We particularly calculate the PSD matrix and discuss its frequency dependence.
Thermal contribution to PSD is estimated by using the FDT, whereas active contribution 
is obtained by assuming that the time correlation of active random forces decays 
exponentially with a characteristic time scale.
The total PSD for a bilayer vesicle is naturally given by the sum of thermal and active contributions. 
We also calculate an apparent response function that can be compared with a recent microrheology 
experiment on RBCs~\cite{Turlier16}. 
We shall argue that the apparent response is enhanced in the low-frequency regime in accordance
with the experimental result.

In Sec.~\ref{sec:relaxation}, we review the relaxation dynamics of a compressible 
bilayer vesicle based on our previous work~\cite{SachinKrishnan16}.
In Sec.~\ref{sec:thermal}, we start with a coupled Langevin equation of a bilayer vesicle and 
discuss the properties of thermal fluctuations by using the FDT. 
The effects of active fluctuations are further discussed in Sec.~\ref{sec:active} by assuming 
exponentially decaying time correlations of active random forces.
The summary and several discussions are finally given in Sec.~\ref{sec:discussion}.

%%%%%%%%%%%%%%%%%%%%
\section{Vesicle relaxation dynamics}
%%%%%%%%%%%%%%%%%%%%
\label{sec:relaxation}

\begin{figure}[b]
\centering
\includegraphics[scale=0.5]{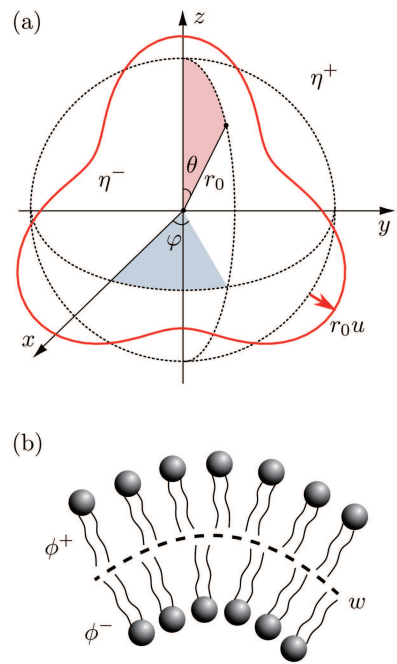}
\caption{
(a) Schematic picture of a vesicle showing the reference sphere (dashed line) of radius $r_0$  
with fluids of viscosity $\eta^-$ inside and $\eta^+$ outside. 
Solid red line represents a deformed vesicle configuration whose shape is parametrized with $u(\theta, \varphi)$. 
(b) Cross-section of a curved region in the bilayer showing the lipid molecules in each monolayer. 
Local densities in the monolayers are indicated as $\phi^+$ and $\phi^-$ which are defined on the 
bilayer midsurface represented by the dashed line. 
The inter-monolayer friction $w$ also acts at the bilayer midsurface.
}
\label{fig:schematic}
\end{figure}

A vesicle is composed of two opposing layers of lipid molecules whose tails meet at the bilayer midsurface.
For a nearly spherical vesicle, the membrane surface can be described by an angle-dependent radius 
field $r (\theta, \varphi)$.
As shown in Fig.~\ref{fig:schematic}, we define a dimensionless deviation in radius $u$ by using the 
radius $r_0$ of a reference sphere as~\cite{Milner87, Faucon89}
\begin{align}
 u (\theta, \varphi) =  \frac{r( \theta, \varphi)}{r_{0}} - 1.
 \label{eq:definition_u}
\end{align}

For bilayer membranes with finite thickness, bending deformation always accompanies stretching 
of one monolayer and compression of the other.
Hence the membrane monolayers should be weakly compressible, and the local lipid density in  
each monolayer is allowed to vary.
Let us denote the number of lipids in the outer and the inner monolayers as $N^+$ and 
$N^-$, respectively.
Since a spherical vesicle is stable only if the outer monolayer has more lipids than the inner one, 
we impose a condition $N^+ > N^-$.
We denote the reference lipid density by $\rho_0 = ( N^+ + N^- ) / ( 2 A_0 )$ where 
$A_0 = 4\pi r_0^2$.
Let $\rho^+ ( \theta, \varphi )$ and $\rho^- ( \theta, \varphi )$ be the variables representing 
the local lipid densities in the two monolayers. 
We then define dimensionless local density deviations in each monolayer as
\begin{align}
\phi^\pm (\theta, \varphi) = \frac{\rho^\pm (\theta, \varphi)}{\rho_0} - 1. 
\label{eq:definition_phi_plus_minus}
\end{align}
Further calculations can be simplified by using the difference and the sum of the densities given by
\begin{align}
\phi^\Delta (\theta, \varphi) = \frac{\phi^+ - \phi^-}{2},~~~~~
\phi^\Sigma (\theta, \varphi) = \frac{\phi^+ + \phi^-}{2},
\label{eq:definition_phi_delta_sigma}
\end{align}
respectively.

In order to describe the configuration of a bilayer vesicle, we make use of the three independent variables 
$u$, $\phi^\Delta$, and $\phi^\Sigma$. 
The free energy of a compressible bilayer vesicle is given by~\cite{Miao02,SachinKrishnan16}
\begin{align}
F &= \int dA \, \Bigg[ \sigma + \frac{\kappa}{2} (2H)^{2} + 2\lambda H \phi^{\Delta} \nonumber \\
&+ k \left[ \left(\phi^\Delta\right)^2  + \left( \phi^\Sigma \right)^2 \right] \Bigg]  + \int dV \, \Delta P,
\label{eq:free_energy}
\end{align}
where $H$ is the local mean curvature that can be expressed in terms of $u$,
$\sigma$ is the membrane surface tension, $\kappa$ is the bending rigidity, $\lambda$ is the 
strength of coupling between local density and curvature, $k$ is the monolayer compressibility, 
and $\Delta P$ is the pressure difference between the inside and the outside of the vesicle.
The first and the second integrals are performed over the area and the volume of the vesicle, 
respectively.

In general, the above three variables are all time-dependent.
Since they are all functions of $\theta$ and $\varphi$, we expand them in terms of surface 
spherical harmonics $Y_{nm}(\theta, \varphi)$ as~\cite{Miao02,SachinKrishnan16}
\begin{align}
u(\theta, \varphi, t) & = \sum_{n=0}^{\infty} \sum_{m=-n}^{n}
u_{nm}(t) Y_{nm}(\theta, \varphi),
\label{eq:u_sh_exp}
\end{align}
and
\begin{align}
\delta \phi^\Delta (\theta, \varphi, t) & = \phi^\Delta - \phi_0^\Delta 
\nonumber \\
& = \sum_{n=2}^{\infty} \sum_{m=-n}^{n}
\phi_{nm}^\Delta (t) Y_{nm}(\theta, \varphi),
\label{eq:phi_delta_sh_exp}
\end{align}
where $\phi_0^\Delta=(N^+-N^-)/(N^++N^-)$.
It can be shown that the mode associated with $\phi^\Sigma$ is decoupled from the 
other two modes and will not be considered hereafter.

The dynamics of a vesicle is affected by the viscosities of the outside and the inside bulk fluids, 
$\eta^{\pm}$~\cite{Komura93,Seki95,Komura96}, and the friction constant between the 
two monolayers, $w$~\cite{Seifert93,Miao02,Fournier15,Okamoto16,SachinKrishnan16}.
We have shown that the relaxation equations for $u$ and $\phi^\Delta$ are given by the set of 
coupled equations~\cite{SachinKrishnan16}
\begin{align}
\frac{\partial}{\partial t} \begin{pmatrix} u_{nm} \\ \phi^\Delta_{nm} \end{pmatrix} 
= - \frac{\bmrm{c} \cdot \bmrm{a}}{\tau} \begin{pmatrix}u_{nm} 
\\ \phi^\Delta_{nm},\end{pmatrix},
\label{eq:dynamical}
\end{align}
for each $(n,m)$-mode.
Here 
\begin{align}
\tau =\frac{\eta^+ r_{0}^{3}}{\kappa}
\label{tau}
\end{align}
is the bending relaxation time and the dimensionless matrices $\bmrm{a}$ and $\bmrm{c}$ are given by
\begin{widetext}
\begin{align}
\renewcommand{\arraystretch}{1.5}
\bmrm{a} = \begin{pmatrix}
(n-1)(n+2)\left[ n(n+1) + \sigma' - 2 \lambda' \phi_0^\Delta + k'\left( \phi^\Delta_0 \right)^2\right] & 4k'\phi^\Delta_0 - 4 \lambda' -\lambda' (n-1)(n+2) \\
4k'\phi^\Delta_0 - 4 \lambda' -\lambda' (n-1)(n+2) & 2k' 
\end{pmatrix},
\label{eq:matrixa}
\end{align}
\begin{align}
\renewcommand{\arraystretch}{1.5}
\bmrm{c} = \frac{n(n+1)}{d} \begin{pmatrix} 4 w' + (2n+1)(1+E) & -(n+2)+(n-1)E \\ -(n+2)+(n-1)E & (n+2)(2n^2-n+2)+(n-1)(2n^2+5n+5)E
\end{pmatrix},
\label{eq:matrixc}
\end{align}
\end{widetext}
where 
\begin{align}
d &= 4 w' \left[(n-1)(2n^2+5n+5)E + (n+2)(2n^2-n+2) \right] \nonumber \\
&+ 2(n-1)(n+1)(2n^2+4n+3)E^2 \nonumber \\
& + \left[8n^2(n+1)^{2}-5\right]E + 2n(n+2)(2n^2+1).
\end{align}
In the above matrices, the dimensionless parameters are defined as 
\begin{align}
& \sigma' = \frac{\sigma r_0^2}{\kappa},~~~
\lambda' = \frac{\lambda r_0}{\kappa},~~~
k' = \frac{kr_0^2}{\kappa},
\\
& w' = \frac{wr_0}{\eta^+},~~~
E = \frac{\eta^-}{\eta^+}.
\end{align}
The symmetric matrix $\bmrm{a}$ represents the inverse susceptibility and is composed 
of the static quantities. 
On the other hand, the matrix $\bmrm{c}$ is constructed by the dynamic quantities and 
corresponds to the kinetic coefficient matrix.
Notice that both $\bmrm{a}$ and $\bmrm{c}$ are symmetric in accordance with the 
Onsager's reciprocal relation~\cite{DoiBook} and depend only on $n$.

In our previous work~\cite{SachinKrishnan16}, we showed that the two relaxation modes 
are coupled to each other as a consequence of the bilayer nature and the spherical structure of the vesicle. 
We investigated the effect of viscosity contrast $E=\eta^-/\eta^+$ on the relaxation rates, and 
found that it linearly shifts the crossover $n$-mode between the bending and the slipping relaxations.
As $E$ is increased, the relaxation rate of the bending mode decreases, while that of the slipping mode
remains almost unaffected.
For parameter values close to the unstable region, some of the relaxation modes are dramatically 
reduced~\cite{SachinKrishnan16}.

%%%%%%%%%%%%%%%%%
\section{Thermal fluctuations}
%%%%%%%%%%%%%%%%%
\label{sec:thermal}

In this section, we discuss thermal fluctuations of a compressible bilayer vesicle.
In the presence of random forces, the coupled memoryless Langevin equation for the two 
variables $u$ and $\phi^\Delta$ can be written as~\cite{DoiBook} 
\begin{align}
\bm{\zeta} \cdot \frac{\partial}{\partial t} \bmrm{x}(t) = 
- \bm{\gamma} \cdot \bmrm{x}(t) + \bm{\xi}(t),
\label{eq:gen_lang_2}
\end{align}
where we have used the vector notation
\begin{align}
\bmrm{x} = r_0 \begin{pmatrix} u_{nm} \\ \phi^\Delta_{nm} \end{pmatrix},
\end{align}
and the matrices
\begin{align}
\bm{\zeta} = \eta^+ r_0 \, \bmrm{c}^{-1},~~~~~
\bm{\gamma} = \frac{\kappa}{r_0^2} \, \bmrm{a},
\label{zeta}
\end{align}
where $\bmrm{c}^{-1}$ represents the inverse matrix.
The vector $\mathbf{x}$ has the dimension of length in order to assign the proper 
dimension for each term in the above Langevin equation.

\begin{table}[bth]
\caption{\label{tab:params}Typical values for the physical quantities.}
\begin{ruledtabular}
\begin{tabular}{ll}
bending modulus: $\kappa$ & $10^{-19}$~J  \\
compressibility: $k$ & $10^{-1}$~J/m$^{2}$ \\
surface tension: $\sigma$ & $10^{-11}$~J/m$^{2}$ \\
outside solvent viscosity: $\eta^{+}$ & $10^{-3}$~J$\cdot$s/m$^{3}$ \\
viscosity contrast: $E =\eta^-/\eta^+$ & $1$, $100$ \\
inter-monolayer friction: $w$ & $10^{9}$~J$\cdot$s/m$^{4}$ \\
vesicle radius: $r_{0}$ & $10^{-5}$~m \\
membrane relaxation time: $\tau=\eta^+ r_{0}^{3}/\kappa$ & $10$~s \\
activity strength: $S_u/(\eta^+ r_0)$ & $80$~$k_{\rm B}T$ \\
activity relaxation time: $\tau_u$ & $10^{-1}$~s \\
\end{tabular}
\end{ruledtabular}
\end{table}

The vector $\bm{\xi}$ represents random forces acting directly on $u$ 
and $\phi^\Delta$: 
\begin{align}
\bm{\xi}(t) = \begin{pmatrix} \xi^{u}_{nm}(t) \\ \xi^\phi_{nm}(t) \end{pmatrix}. 
\end{align}
In thermal equilibrium, these forces have zero mean and obey the following fluctuation 
dissipation theorem (FDT)~\cite{DoiBook}:
\begin{align}
& \langle \bm{\xi}(t) \rangle  = 0,
\label{eq:noisemean}
\\
& \langle \bm{\xi} (t_1) \bm{\xi}^{\textrm T} (t_2) \rangle = 
2 k_{\textrm B} T \bm{\zeta} \delta (t_1 - t_2),
\label{eq:noise}
\end{align}
where ``T" indicates the transpose operator, $k_{\rm B}$ is the Boltzmann constant, 
and $T$ is the temperature.
Notice that the friction matrix $\bm{\zeta}$ (or $\bmrm{c}^{-1}$) contains 
off-diagonal elements, implying that thermal fluctuations of the two fields $u$ and 
$\phi^{\Delta}$ are correlated with each other. 
This is a unique feature of spherically closed vesicles because such a coupling effect does not 
exist for planar bilayer membranes~\cite{SachinKrishnan16}.

\begin{figure}[tbh]
\centering
\includegraphics[scale=0.4]{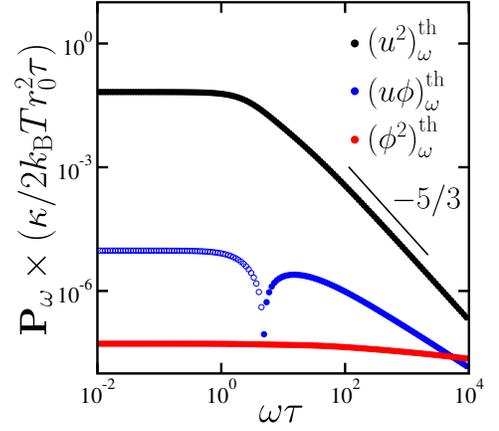}
\caption{
Dimensionless components of thermal power spectral density (PSD) matrix as a function of 
dimensionless frequency $\omega \tau$.
The characteristic bending relaxation time is $\tau=\eta^+ r_{0}^{3}/\kappa$ which is 
chosen here as $\tau=10$~s (see Table I).
The dimensionless parameter values are 
$\sigma' = 10^{-2}$, 
$\lambda' = 10^{4}$,
$k' = 10^8$,  
$w' = 10^7$, 
$\phi_0^\Delta = 3 \times 10^{-4}$, $E = 1$ 
(these values can be obtained from Table I).
Three independent components are plotted;
$(u^2)_\omega^{\rm th}$ (black), 
$(u\phi)_\omega^{\rm th}$ (blue), 
and $(\phi^2)_\omega^{\rm th}$ (red).
The component $(u\phi)_\omega^{\rm th}$ becomes negative for small frequencies 
as represented by open blue circles.
The slope of $-5/3$ is drawn for comparison.
}
\label{fig:thermal_psd}
\end{figure}

\begin{figure}[tbh]
\centering
\includegraphics[scale=0.4]{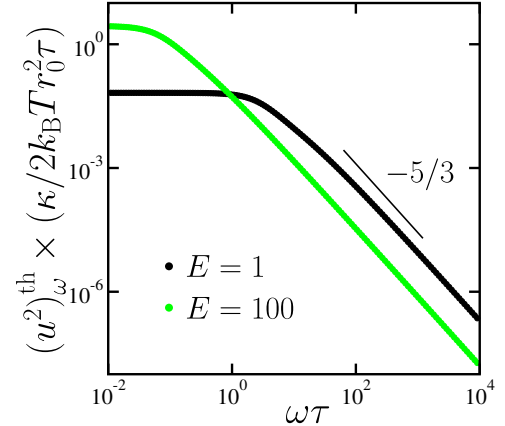}
\caption{
Dimensionless thermal power spectral density $(u^2)_\omega^{\rm th}$ as a function of 
dimensionless frequency $\omega \tau$ for $E=1$ (black) and $E=100$ (green). 
The other parameters are the same with those in Fig.~\ref{fig:thermal_psd}.
The slope of $-5/3$ is drawn for comparison.
}
\label{fig:viscont_psd}
\end{figure}

The Fourier components of the $2\times2$ correlation function matrix can be obtained by
\begin{align}
\left( \bmrm{x} \bmrm{x}^{\textrm T} \right)_\omega = 
\int_{-\infty}^{\infty} dt \,  
\langle \bmrm{x}(t) \bmrm{x}^{\textrm T} (0) \rangle e^{i\omega t},
\label{eq:x2omega}
\end{align}
where $\omega$ is the frequency. 
With the use of Eqs.~(\ref{eq:noisemean}) and (\ref{eq:noise}), the thermal contribution 
to the correlation function matrix is given by~\cite{Landau69}
\begin{align}
 (\bmrm{x} \bmrm{x}^{\textrm T} )_\omega^{\rm th}  
= 2 k_{\rm B}T   \left(\bm{\gamma} - i \omega \bm{\zeta} \right)^{-1} 
\cdot \bm{\zeta} 
\cdot \left(\bm{\gamma} + i \omega \bm{\zeta} \right)^{-1},
\label{eq:fluc}
\end{align}
whose components depend on $n$.
Then the power spectral density (PSD) matrix is obtained by summing over all the 
$n$-modes as 
\begin{align}
\bmrm{P}_{\omega}^{\rm th} & =
\begin{pmatrix}
(u^2)_\omega^{\rm th} &  (u\phi)_\omega^{\rm th} \\
(\phi u)_\omega^{\rm th} & (\phi^2)_\omega^{\rm th}  
\end{pmatrix} 
\nonumber \\ 
& = \sum_{n=2}^{n_{\rm max}} (2n+1) (\bmrm{x} \bmrm{x}^{\textrm T})_\omega^{\rm th}.
\label{eq:thermal_psd}
\end{align}
Notice that the components of the correlation function matrix do not depend on $m$ but are
$(2n+1)$-fold degenerated.
In the following numerical estimate, we set $n_{\rm max}=1000$ which covers sufficient 
length scales for the present analysis.

In Fig.~\ref{fig:thermal_psd}, we plot the frequency dependence of each component of the 
numerically calculated PSD matrix in Eq.~(\ref{eq:thermal_psd}) for a vesicle of size $r_0 = 10$~$\mu$m.
The other parameter values and the relevant quantities are summarized in Table~\ref{tab:params}.
The three curves correspond to the three components of the symmetric PSD matrix 
$(u^2)_\omega^{\rm th}$ (black), $(u\phi)_\omega^{\rm th}$ (blue), and $(\phi^2)_\omega^{\rm th}$ (red).
Within the present scaled unit, $(u^2)_\omega^{\rm th}$ is much larger than 
$(\phi^2)_\omega^{\rm th}$ and $(u\phi)_\omega^{\rm th}$.
Hence the density field fluctuations are relatively small. 
The PSD component $(u\phi)_\omega^{\rm th}$
lies in between $(u^2)_\omega^{\rm th}$ and 
$(\phi^2)_\omega^{\rm th}$, and takes negative values in the small frequency region
(shown by open blue circles).

In general, all the PSD components are almost constant in the low-frequency regime
and decay in the high-frequency regime.
The important crossover frequency $\omega^{\ast}$ is set by the membrane relaxation time 
$\tau=\eta^+ r_{0}^{3}/\kappa$ and is given by $\omega^{\ast} \approx 1/\tau$. 
For the parameter values in Table I, the characteristic relaxation time is chosen as $\tau=10$~s
and $\omega^{\ast}=0.1$~s$^{-1}$. 
In the low frequency regime of $\omega \tau \ll 1$ (or $\omega \ll \omega^{\ast}$), all 
the PSD components become independent of the frequency, which is a characteristic 
feature of a vesicle with a finite size.
In the high frequency limit of $\omega \tau \gg 1$ (or $\omega \gg \omega^{\ast}$), the 
PSD $(u^2)_\omega^{\rm th}$ shows a power-law decay with an exponent $-1.60$ which 
is close to the exponent $-5/3$ obtained by Zilman and Granek for a flat membrane without 
any bilayer structure~\cite{Zilman96, Zilman02}.  
We also note that the frequency dependence of $(\phi^2)_\omega^{\rm th}$ is very 
weak in the entire region.
Hence, we shall mainly focus on the behavior of displacement fluctuations $(u^2)_\omega$ 
in the following discussion.

To see the effect of viscosity contrast between the inside and the outside of a vesicle,
we plot in Fig.~\ref{fig:viscont_psd} the PSD component $(u^2)_\omega^{\rm th}$ 
for $E=\eta^{-}/\eta^{+}=1$ (the same as in Fig.~\ref{fig:thermal_psd}) and $E=100$
(large inside viscosity~\cite{Fujiwara14}). 
When the inside viscosity becomes larger, the crossover frequency $\omega^{\ast}$ 
becomes smaller and the fluctuation amplitude is enhanced in the low-frequency region.
This is because the effective viscosity for $E\neq 1$ is essentially given by the combination 
of the two viscosities such that $\eta_{\rm eff} \approx \eta^{+}+\eta^{-}$.
In the high frequency regime, on the other hand, the PSD for $E=100$ is smaller than that 
of $E=1$ because the displacement fluctuations are suppressed when the inner viscosity 
is larger.
However, the power-law behavior in the large frequency limit ($\omega \gg \omega^{\ast}$) of the PSD with an exponent $-1.6 \approx -5/3$ 
is the same for the two cases of $E=1$ and $100$.

%%%%%%%%%%%%%%%%
\section{Active fluctuations}
%%%%%%%%%%%%%%%%
\label{sec:active}

\begin{figure}[tbh]
\centering
\includegraphics[scale=0.4]{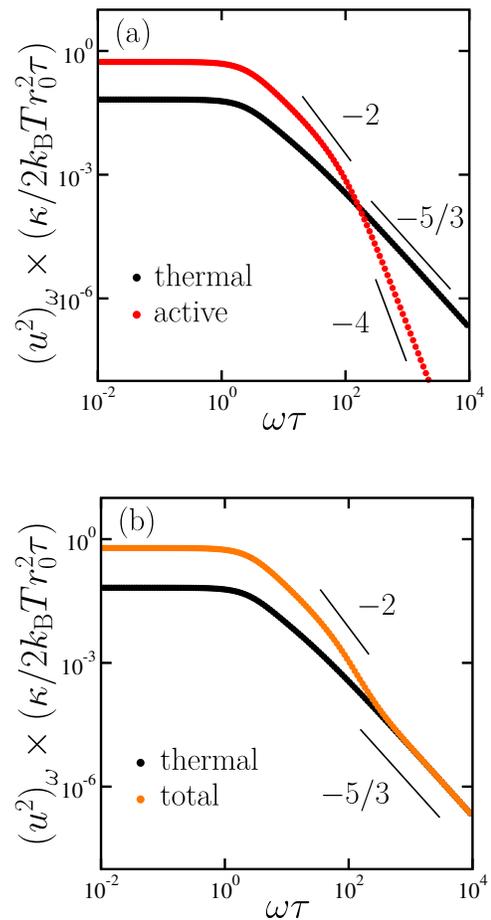}
\caption{
(a) Dimensionless thermal power spectral density $(u^2)_\omega^{\rm th}$ (black) and 
active power spectral density $(u^2)_\omega^{\rm ac}$ (red) as a function 
of dimensionless frequency $\omega \tau$.
The parameters for the active contribution are 
$S_u/(\eta^+ r_0) = 80$~$k_{\rm B}T$, 
$\tau_u/\tau = 10^{-2}$, 
$S_\phi= 0$
(see Table I).
The other parameters are the same with those in Fig.~\ref{fig:thermal_psd}.
(b) Dimensionless thermal power spectral density $(u^2)_\omega^{\rm th}$ (black) and 
total power spectral density $(u^2)_\omega=(u^2)_\omega^{\rm th}+(u^2)_\omega^{\rm ac}$ 
(orange) as a function of dimensionless frequency $\omega \tau$.
The slopes of $-5/3$, $-2$, and $-4$ are drawn for comparison.
}
\label{fig:activity_psd}
\end{figure}

In this section, we discuss the influence of nonequilibrium random forces on the 
power spectrum of fluctuations.
In real cells, these active forces are generated as a consequence of metabolic 
processes such as the pumping action of ion channels or the growth of cytoskeletal 
networks~\cite{Turlier16}.
The modified coupled Langevin equation in the presence of active random forces is given by 
\begin{align}
\bm{\zeta} \cdot \frac{\partial }{\partial t} \bmrm{x}(t) = 
- \bm{\gamma} \cdot \bmrm{x} (t) + \bm{\xi} (t) + \bm{\mu} (t),
\label{eq:gen_lang_3}
\end{align}
where we have added the following vector of active random forces to Eq.~(\ref{eq:gen_lang_2}); 
\begin{align}
\bm{\mu} (t) = \begin{pmatrix} \mu^u_{nm} (t) \\ \mu^\phi_{nm} (t) \end{pmatrix}.
\end{align}
We assume that these random forces have zero mean and their cross correlations vanish, i.e.,
\begin{align}
& \langle \bm{\mu} (t) \rangle  = 0,
\\
& \langle \mu^u_{nm} (t_1) \mu^{\phi}_{n'm'} (t_2) \rangle = 0.
\end{align}

\begin{figure}[tbh]
\centering
\includegraphics[scale=0.4]{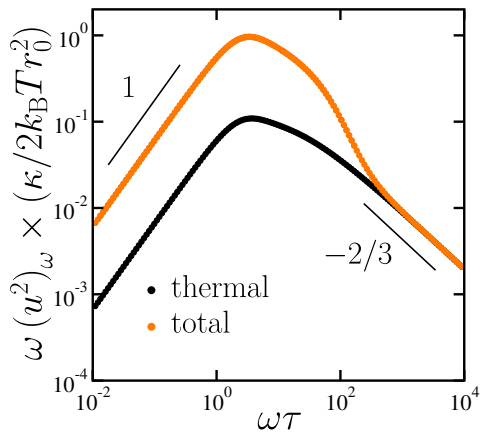}
\caption{
Dimensionless thermal response function $\omega (u^2)_\omega^{\rm th}/(2 k_{\rm B}T)$ (black) 
and total apparent response function 
$\omega (u^2)_\omega/(2 k_{\rm B}T)
=[(u^2)_\omega^{\rm th}+(u^2)_\omega^{\rm ac}]/(2 k_{\rm B}T)$ 
(orange) as a function of dimensionless frequency $\omega \tau$.
The other parameters are the same with those in Fig.~\ref{fig:activity_psd}.
The slopes of $-2/3$ and $1$ are drawn for comparison.
}
\label{fig:response}
\end{figure}

In the above, active fluctuations operate directly on the displacement and the density fields.
They can be used to model any of the several metabolic processes that modify the membrane dynamics.
For instance, nonequilibrium fluctuations acting on the displacement field mimic processes that affect the 
membrane dynamics in the radial direction such as pumping of ion channels or cytoskeletal growth.
On the other hand, fluctuations acting on the density field can represent the action of flippase 
proteins which consume ATP and facilitate exchange of lipid molecules between the two monolayers.

In general, the membrane dynamics can be affected by several active processes with 
different characteristic timescales.
For simplicity, however, we consider the case when there is only one dominant active 
process characterized by a single timescale for all the modes.
Then the correlations of active random forces are assumed to have the forms 
\begin{align}
\langle \mu^u_{nm} (t_1) \mu^{u}_{n'm'} (t_2) \rangle 
&=\frac{S_u}{2\tau_u} \, e^{-|t_1 - t_2|/\tau_u} \delta_{nn'} \delta_{mm'}, 
\\
\langle \mu^\phi_{nm} (t_1) \mu^{\phi}_{n'm'} (t_2) \rangle 
&= \frac{S_{\phi}}{2\tau_{\phi}} \, e^{-|t_1 - t_2|/\tau_{\phi}} \delta_{nn'} \delta_{mm'}, 
\end{align}
where $\tau_u$ and $\tau_{\phi}$ are the characteristic time scales, while $S_u$ and 
$S_{\phi}$ denote the strengths of the activity.
Here we also assume that these quantities depend neither on $n$ nor on $m$.

The Fourier transform of the above active force correlations can be expressed in the matrix form as 
\begin{align}
(\bm{\mu} \bm{\mu}^{\textrm T})_\omega
= \begin{pmatrix}  \dfrac{S_u}{1 + (\omega \tau_u )^2}  & 0 \\ 0 & \dfrac{S_{\phi } }{1 + (\omega \tau_{\phi} )^2}\end{pmatrix},
\end{align}
where the diagonal elements are Lorentzian functions.
Using this expression, we can write the active contribution to the correlation function matrix as 
\begin{align}
 (\bmrm{x} \bmrm{x}^{\textrm T} )_\omega^{\rm ac}  = 
\left(\bm{\gamma} - i \omega \bm{\zeta} \right)^{-1} 
\cdot (\bm{\mu} \bm{\mu}^{\textrm T})_\omega  
\cdot \left(\bm{\gamma} + i \omega \bm{\zeta} \right)^{-1},
\end{align}
whose components depend only on $n$.
Then the active PSD matrix can be calculated by  
\begin{align}
\bmrm{P}_{\omega}^{\rm ac} & =
\begin{pmatrix}
(u^2)_\omega^{\rm ac} &  0 \\
0 & (\phi^2)_\omega^{\rm ac}  
\end{pmatrix} 
\nonumber \\ 
& =\sum_{n=2}^{n_{\rm max}} (2n+1) (\bmrm{x} \bmrm{x}^{\textrm T})_\omega^{\rm ac},
\label{eq:active_psd}
\end{align}
similar to the thermal contribution in Eq.~(\ref{eq:thermal_psd}).
Finally, the total PSD is given by the sum of thermal and active contributions;
\begin{align}
\bmrm{P}_{\omega} = \bmrm{P}_{\omega}^{\rm th} + 
\bmrm{P}_{\omega}^{\rm ac}.
\label{Psum}
\end{align}
Here we have implicitly assumed that thermal and active random forces are not 
correlated with each other.

In Fig.~\ref{fig:activity_psd}(a), we plot both the thermal PSD component 
$(u^2)_\omega^{\rm th}$ (black, the same as in Fig.~\ref{fig:thermal_psd}) 
and the active PSD component $(u^2)_\omega^{\rm ac}$ (red) as a function of $\omega \tau$ 
when $S_u/(\eta^+ r_0) = 80$~$k_{\rm B}T$, $\tau_u/\tau = 10^{-2}$, and 
$S_\phi= 0$ neglecting active fluctuations of the density field $\phi^{\Delta}$.
Since the membrane relaxation time is chosen as $\tau=10$~s for a vesicle of size 
$r_0 = 10$~$\mu$m, the above parameter means that the activity time scale is 
roughly $\tau_u = 10^{-1}$~s~\cite{Turlier16}.
The active contribution decays as $\sim \omega^{-2}$ in the intermediate frequency regime, 
$\tau^{-1} \ll \omega \ll \tau_u^{-1}$, and further decays as $\sim \omega^{-4}$  in the high 
frequency region, $\omega \tau_u \gg 1$. 
In Fig.~\ref{fig:activity_psd}(b), we plot the total PSD component $(u^2)_\omega$ (orange)
given by Eq.~(\ref{Psum}) as well as the thermal PSD (black, the same as in (a))
in order to compare between the equilibrium and nonequilibrium situations.
We note that active contribution becomes important in the low frequency region
$\omega \tau_u \ll 1$ and enhances the total fluctuations in the nonequilibrium 
active case.

To present our result in a different way, we plot in Fig.~\ref{fig:response} an apparent 
total response function defined by $\omega \bmrm{P}_{\omega}/(2 k_{\rm B}T)$ (orange) 
as a function of $\omega \tau$.
We use the same parameters as in Fig.~\ref{fig:activity_psd}.
In thermal equilibrium, the response function 
$\omega \bmrm{P}_{\omega}^{\rm th}/(2 k_{\rm B}T)$ (black) should satisfy the 
relation~\cite{Landau69} 
\begin{align}
\frac{\omega \bmrm{P}_{\omega}^{\rm th}}{2k_{\rm B}T} =\bm{\chi}'' (\omega),
\label{eq:fdt}
\end{align}
where $\bm{\chi}'' (\omega)$ is the imaginary part of the complex response function 
matrix given by 
\begin{align}
\bm{\chi}(\omega) = \sum_{n=2}^{n_{\rm max}} (2n + 1) 
\left(\bm{\gamma} - i \omega \bm{\zeta} \right)^{-1}.
\end{align}
Here the response function matrix can be obtained from Eq.~(\ref{eq:gen_lang_3})
when both thermal and active fluctuations are absent, i.e., $\bm{\xi}=\bm{\mu}=0$.
In nonequilibrium case, the apparent response function 
$\omega \bmrm{P}_{\omega}/(2 k_{\rm B}T)$  (orange) is enhanced.
Moreover, it increases linearly with $\omega$ in the low frequency regime, while it exhibits 
a power-law decay with an exponent $-2/3$ in the high frequency regime. 
As we shall discuss in the next section, these results are consistent with the experimental 
findings by Turlier \textit{et al.}~\cite{Turlier16} who used a more detailed model to take 
into account the activity.

Generally speaking,  Eq.~(\ref{eq:fdt}) is an alternative expression of the FDT~\cite{Landau69}.
It states that the thermal PSD matrix $\bmrm{P}^{\rm th}_{\omega}$ and the imaginary part of 
the response function $\bm{\chi}''(\omega)$ are intimately related in equilibrium.
In the presence of active fluctuations, the total PSD of fluctuations includes both thermal and 
active contributions $\bmrm{P}^{\rm th}_{\omega} + \bmrm{P}^{\rm ac}_{\omega}$ 
as in Eq.~(\ref{Psum}), whereas an experimentally measured response function should not be 
altered even in nonequilibrium situations.
Therefore, the FDT is inevitably violated in the presence of active fluctuations, and the extent 
of violation is generally characterized by $\bmrm{P}^{\rm ac}_{\omega}$.

%%%%%%%%%%%%%%%%%%%
\section{Summary and discussions}
%%%%%%%%%%%%%%%%%%%
\label{sec:discussion}

In this paper, we have discussed the statistical properties of thermal and active fluctuations 
of a compressible bilayer vesicle.
We have analyzed the coupled Langevin equation for the membrane deformation and the 
lipid density variables.
In particular, we have calculated the PSD matrix of these fluctuations and discussed the 
frequency dependence of each component.
Thermal contribution to PSD is estimated by means of the FDT, whereas active contribution 
is obtained by assuming that the time correlation functions of active random forces decay 
exponentially with a characteristic time scale.
We have assumed that the total PSD is given by the sum of thermal and active contributions. 
Our results show that the nonequilibrium active contribution affects fluctuations below a 
characteristic frequency set by the activity timescale. 
To compare with the recent microrheology experiment on RBCs~\cite{Turlier16}, we have 
further obtained an apparent response function which is enhanced in the low-frequency regime.

Recently, Turlier \textit{et al.}\ reported a FDT violation and argued that the cell membrane 
fluctuations in healthy RBCs are actively driven~\cite{Turlier16}. 
The experimentally measured fluctuations and response function were in agreement with their 
theoretical calculations obtained from the spectrin network model of active membranes.
They showed that the timescale of switching between active and inactive states of the 
phosphorylation sites on the spectrin network is roughly $0.15$~s~\cite{Turlier16}.
In order to make a comparison with their experiment, we have set the activity timescale 
$\tau_u = 0.1$~s to calculate the apparent response function in Fig.~\ref{fig:response}.
We have shown that the enhanced response function increases linearly at low frequencies 
$\omega \tau \ll 1$ and decays as $\sim \omega^{-2/3}$ at very high frequencies 
$\omega \tau_{u} \gg 1$.
These results are in accordance with the experimental result by Turlier 
\textit{et al.}~\cite{Turlier16}.

Most of the existing works on membrane fluctuations describe only the dynamics of the 
displacement field $u$.
In this work, we have taken into account the compressibility of the membrane and have 
also discussed the dynamics of the lipid density field $\phi^{\Delta}$.
In order to write down the coupled Langevin equation in Eq.~(\ref{eq:gen_lang_2}), we have used 
the fact that the dynamics of these two variables is expressed by the product of 
the two symmetric matrices, i.e., the inverse susceptibility matrix $\bmrm{a}$ in 
Eq.~(\ref{eq:matrixa}) and the kinetic coefficient matrix $\bmrm{c}$ in Eq.~(\ref{eq:matrixc}). 
The obtained PSD matrix characterizes the frequency dependencies of fluctuation amplitudes 
for the membrane displacement and the lipid density.
Although some works have reported the effects of bilayer nature on the static fluctuation 
spectrum of the membrane displacement~\cite{Watson10,Mell15}, we are not aware of any 
experimental work where the lipid density fluctuation is investigated explicitly.

The thermal PSD component $(u^2)^{\rm th}_{\omega}$ is related to the Fourier transform of 
the mean squared displacement (MSD) of a tagged membrane segment.
This is because the MSD and the displacement correlation function are related by 
$\langle [u(t)-u(0)]^2 \rangle = 2[\langle u^2(0) \rangle - \langle u(t) u(0) \rangle]$ and the time 
dependencies of the MSD and the correlation function are the same except for the numerical 
prefactor (as long as the equal time correlation function $\langle u^2(0) \rangle$ is 
finite)~\cite{DoiBook}.
For a flat membrane, Zilman and Granek predicted that a membrane segment undergoes anomalous
out-of-plane diffusion for which the MSD increases as $\sim t^{2/3}$~\cite{Zilman96,Zilman02}. 
In the frequency domain, this behavior is consistent with our result of 
$(u^2)^{\rm th}_{\omega} \sim \omega^{-5/3}$ as shown in Fig.~\ref{fig:thermal_psd}.
Therefore, we confirm that the scaling relation for a flat membrane also holds for a 
compressible bilayer vesicle. 
Moreover, we find that this scaling behavior is not affected by the viscosity contrast $E$
as shown in Fig.~\ref{fig:viscont_psd}.
The thermal PSD component $(u^2)^{\rm th}_{\omega}$ is independent of $\omega$ at low 
frequencies ($\omega \tau \ll 1$ or $\omega \ll \omega^{\ast}$) because the slowest bending 
relaxation time $\tau$ is set by the finite size of a vesicle.

In our theory, both the inertia of the membrane segment and the inertia of the surrounding fluid 
are neglected when we calculate the hydrodynamic response of a bilayer vesicle. 
Within this approximation, we have employed the standard memoryless Langevin equation in 
the overdamped form. 
On the other hand, it is known that the Brownian motion of a spherical particle in an incompressible 
fluid should be described by the generalized Langevin equation~\cite{KuboBook} with the 
Boussinesq force~\cite{LandauFluid}. 
For a spherical particle in a 3D fluid, the correction to the Stokes friction due to the fluid inertia 
(called the Basset force) leads to a long-time tail behavior of the velocity autocorrelation function 
(typically $\sim t^{-3/2}$)~\cite{Franosch11,Kheifets14}.

It is beyond the scope of the present paper to calculate the hydrodynamic memory effect of fluid inertia
on the membrane friction matrix $\bm{\zeta}$ [see Eq.~(\ref{zeta})]. 
However, we can roughly estimate the crossover frequency $\omega^{\ast\ast}$ above which the 
fluid inertia plays a role. 
The minimum of the crossover frequency is roughly given by $\omega^{\ast\ast} = \eta^+/(\rho r_0^2)$,
where $\eta^+$ is the fluid viscosity, $\rho$ is the fluid density, and $r_0$ corresponds to the vesicle 
radius which is the largest size in the problem. 
With the value $\rho \approx 10^3$~kg/m$^3$ for water and the other values quoted in Table I, 
we obtain $\omega^{\ast\ast} \approx 10^4$~s$^{-1}$ and $\omega^{\ast\ast} \tau \approx 10^5$, 
where $\tau \approx 10$~s is the bending relaxation time of the vesicle [see Eq.~(\ref{tau})]. 
Generally, inertial effects of the fluid become strong at frequencies higher than $\omega^{\ast\ast}$. 
In other words, effects of fluid inertia and stress propagation are negligible for frequencies 
$\omega \ll \omega^{\ast\ast}$.
Furthermore, an analogous crossover frequency for a membrane segment (rather than a whole vesicle) 
can be even larger because its size is much smaller than the vesicle size $r_0$.

The frequency range considered in Figs.~\ref{fig:thermal_psd}--\ref{fig:response} 
is $10^{-2} \le \omega \tau \le 10^4$ for which we are safely allowed to neglect inertial effects 
of the surrounding bulk fluid. 
Although our calculation relies on the Langevin equation without any hydrodynamic memory effect, 
the ``large frequency'' regime in the graphs ($1 \ll \omega \tau \le 10^4$) corresponds to the 
``small frequency'' regime when compared with the above estimated crossover frequency 
$\omega^{\ast\ast} \tau \approx 10^5$
related to the fluid inertia. 
This argument justifies the usage of the standard Langevin equation and allows us to discuss the 
long time behavior of membrane fluctuations. 
We also comment that the frequency range $10^{-3}$--$10^3$~s$^{-1}$ plotted in the graphs 
are easily accessible by the current experimental techniques.

Here we shall briefly mention that both the inertia of the fluid membrane and the surrounding fluid were
considered by Camley and Brown~\cite{Camley11} and by 
Komura \textit{et al}.~\cite{Seki93,Komura12} for an infinitely large and flat membrane without any 
out-of-plane deformations. 
Especially, Camley and Brown showed in the former paper that fluid inertia creates a long-time tail in the 
membrane velocity correlation function. 
In a viscous membrane, this tail crosses over from $\sim t^{-1}$ (as in a 2D fluid) at intermediate times to 
$\sim t^{-3/2}$ (as in a 3D fluid) at long times. 
An analogous study for membranes undergoing out-of-plane fluctuations can be interesting.
However, it was also pointed out that a long-time tail behavior in a membrane becomes important 
only when the correlation function reaches $10^{-6}$ of its initial value, which 
would be effectively unobservable for vesicles~\cite{Camley11}.

On the other hand, the memory effect due to the viscoelasticity of the surrounding media is worth considering. 
When the outer fluid is viscoelastic, the constant viscosity $\eta$ in the Stokes equation should be replaced 
by a frequency dependent viscosity $\eta(\omega)$ in Eq.~(\ref{zeta}) according to the correspondence 
principle of linear viscoelasticity~\cite{Grimm11,FurstBook}. 
Granek used a generalized Langevin equation with a memory kernel to calculate the membrane 
MSD~\cite{Granek11}.
He also showed that, if the complex shear viscosity of the fluid is described by a power-law behavior 
$\eta(\omega) \sim (i \omega) ^{\alpha-1}$, a tagged membrane segment would exhibit an 
anomalous subdiffusive behavior and its MSD should grow as $\sim t^{2\alpha/3}$.
For $\alpha = 1$, we  recover the scaling behavior for a purely viscous medium as described before.
Thus, we expect the PSD to scale as $\sim \omega ^ {-(2 \alpha +3) / 3}$ when the vesicle is surrounded 
by a viscoelastic medium which can mediate mechanical memory effects.

Finally, it is worth mentioning that we have chosen simple forms for active noise correlations that 
capture only limited features of active processes such as their timescale and strength.
As an extension of the present work, we shall consider a coupling effect between active fluctuations 
and membrane local curvature.
Such a coupling can be taken into account through a wavenumber dependent activity 
strength~\cite{Gov04}, and it will alter the frequency dependence of active fluctuations.
In general, the frequency dependency of PSD can be used to characterize and distinguish 
between different active noise sources in experiments.

%%%%%%%%%%
\acknowledgments
%%%%%%%%%%

T.V.S.K.\ acknowledges Tokyo Metropolitan University for the support provided through the 
co-tutorial program.
S.K.\ and R.O.\ acknowledge support by Grant-in-Aid for Scientific Research on
Innovative Areas ``\textit{Fluctuation and Structure}" (Grant No.\ 25103010) from the Ministry
of Education, Culture, Sports, Science, and Technology (MEXT) of Japan and by
Grant-in-Aid for Scientific Research (C) (Grant No.\ 15K05250) from the Japan Society for the 
Promotion of Science (JSPS).

%%%%%%%%%%%%%%%


\begin{thebibliography}{99}
%%%%%%%%%%%%%%%

\bibitem{SafranBook}
S. A. Safran, 
\textit{Statistical Thermodynamics of Surfaces, Interfaces, and Membranes} 
(Addison-Wesley, Reading, MA, 1994).

\bibitem{Rodriguez-Garcia09}
R. Rodr\'{i}guez-Garc\'{i}a, L. R. Arriaga, M. Mell, L. H. Moleiro, I. L\'{o}pez-Montero, and F. Monroy,
Phys. Rev. Lett. \textbf{102}, 128101 (2009).

\bibitem{Boss12}
D. Boss, A. Hoffmann, B. Rappaz, C. Depeursinge, P. J. Magistretti, D. V. de Ville, and P. Marquet,
PLoS ONE \textbf{7}, e40667 (2012).

\bibitem{Faucon89}
J. F. Faucon, M. D. Mitov, P. M\'{e}l\'{e}ard, I. Bivas, and P. Bothorel,
J. Phys. (Paris) \textbf{50}, 2389 (1989).

\bibitem{Milner87}
S. T. Milner and S. A. Safran,
Phys. Rev. A \textbf{36}, 4371 (1987).

\bibitem{Brochard75}
F. Brochard and J. F. Lennon,
J. Phys. France \textbf{36}, 1035 (1975).

\bibitem{Betz09}
T. Betz, M. Lenz, J.-F. Joanny, and C. Sykes,
Proc. Natl Acad. Sci. USA \textbf{106}, 15320 (2009).

\bibitem{Park10}
Y. Park, C. A. Best, T. Auth, N. S. Gov, S. A. Safran, G. Popescu, S. Suresh, and M. S. Feld,
Proc. Natl Acad. Sci. USA \textbf{107}, 1289 (2010).

\bibitem{Turlier16}
H. Turlier, D. A. Fedosov, B. Audoly, T. Auth, N. S. Gov, C. Sykes, J.-F. Joanny, G. Gompper, and T. Betz,
Nature Physics \textbf{12}, 513 (2016).

\bibitem{Biswas17}
A. Biswas, A. Alex, and B. Sinha,
Biophys. J. \textbf{113}, 1768 (2017).

\bibitem{Almendro-Vedia17}
V. G. Almendro-Vedia, P. Natale, M. Mell, S. Bonneau, F. Monroy, F. Joubert, and I. L\'{o}pez-Montero,
Proc. Natl Acad. Sci. \textbf{114}, 11291 (2017).

\bibitem{Fenz17}
S. F. Fenz, T. Bihr, D. Schmidt, R. Merkel, U. Seifert, K. Sengupta, and A. Smith,
Nature Physics \textbf{13}, 906 (2017). 

\bibitem{Prost96}
J. Prost and R. Bruinsma,
Europhys. Lett. \textbf{33}, 321 (1996).

\bibitem{Ramaswamy00}
S. Ramaswamy, J. Toner, and J. Prost,
Phys. Rev. Lett. \textbf{84}, 3494 (2000).

\bibitem{Lacoste05}
D. Lacoste and A. W. C. Lau,
Europhys. Lett. \textbf{70} (3), 418 (2005).

\bibitem{Komura15}
S. Komura, K. Yasuda, and R. Okamoto, 
J. Phys.: Condens. Matter \textbf{27}, 432001 (2015).

\bibitem{Gov04}
N. Gov,
Phys. Rev. Lett. \textbf{93}, 268104 (2004).

\bibitem{Gov05}
N. S. Gov and S. A. Safran,
Biophys. J. \textbf{88}, 1859 (2005).

\bibitem{Gov07}
N. S. Gov,
Phys. Rev. E \textbf{75}, 011921 (2007).

\bibitem{Yasuda16}
K. Yasuda, S. Komura, and R. Okamoto,
Phys. Rev. E \textbf{93}, 052407 (2016).

\bibitem{Seifert93}
U. Seifert and S. A. Langer,
Europhys. Lett. \textbf{23}, 71 (1993).

\bibitem{Miao02}
L. Miao, M. A. Lomholt, and J. Kleis,
Eur. Phys. J. E \textbf{9}, 143 (2002).

\bibitem{Fournier15}
J.-B. Fournier, 
Int. J. Nonlinear Mech. \textbf{75}, 67 (2015).

\bibitem{Okamoto16}
R. Okamoto, Y. Kanemori, S. Komura, and J.-B. Fournier, 
Eur. Phys. J. E \textbf{39}, 52 (2016).

\bibitem{SachinKrishnan16}
T. V. Sachin Krishnan, R. Okamoto, and S. Komura,
Phys. Rev. E \textbf{94}, 062414 (2016).

\bibitem{DoiBook}
M. Doi, 
\textit{Soft Matter Physics} 
(Oxford University, Oxford, 2013).

\bibitem{Okamoto17}
R. Okamoto, S. Komura, and J.-B. Fournier,
Phys. Rev. E \textbf{96}, 012416 (2017).

\bibitem{Komura93}
S. Komura and K. Seki,
Physica  A \textbf{192}, 27 (1993).

\bibitem{Seki95}
K. Seki and S. Komura,
Physica  A \textbf{219}, 253 (1995).

\bibitem{Komura96}
S. Komura,
in \textit{Vesicles}, edited by M. Rosoff (Marcel Dekker, Inc., New York, 1996), p. 198.

\bibitem{Landau69}
L. D. Landau and E. M. Lifshitz,
\textit{Statistical Physics} 
(Pergamon Press, Oxford, 1980).

\bibitem{Zilman96}
A. G. Zilman and R. Granek,
Phys. Rev. Lett. \textbf{77}, 4788 (1996).

\bibitem{Zilman02}
A. G. Zilman and R. Granek,
Chem. Phys. \textbf{284}, 195 (2002).

\bibitem{Fujiwara14}
K. Fujiwara and M. Yanagisawa,
ACS Synth. Biol. \textbf{3}, 870 (2014).
  
\bibitem{Watson10}
M. C. Watson and F. L. H. Brown,
Biophys. J \textbf{98}, L09 (2010).

\bibitem{Mell15}
M. Mell, L. H. Moleiro, Y. Hertle, I. L\'{o}pez-Montero, F. J. Cao, P. Fouquet, T. Hellweg, and F. Monroy,
Chem. Phys. Lipids \textbf{185}, 61 (2015).

\bibitem{KuboBook}
R. Kubo, M. Toda, and N. Hashitsume, 
\textit {Statistical Physics II}
(Springer-Verlag, New York, 1991).

\bibitem{LandauFluid}
L. D. Landau and E. M. Lifshitz,
\textit{Fluid Mechanics} 
(Pergamon Press, Oxford, 1987).

\bibitem{Franosch11}
T. Franosch, M. Grimm, M. Belushkin, F. M. Mor, G. Foffi, L. Forr\'{o}, and S. Jeney, 
Nature \textbf{478}, 85 (2011). 

\bibitem{Kheifets14}
S. Kheifets, A. Simha, K. Melin, T. Li, and M. G. Raizen,
Science \textbf{343}, 1493 (2014).

\bibitem{Camley11}
B. A. Camley and F. L. H. Brown,
Phys. Rev. E \textbf{84}, 021904 (2011). 

\bibitem{Seki93}
K. Seki and S. Komura,
Phys. Rev. E \textbf{47}, 2377 (1993).

\bibitem{Komura12}
S. Komura, S. Ramachandran, and K. Seki,
EPL \textbf{97}, 68007 (2012).

\bibitem{Grimm11}
M. Grimm, S. Jeney, and T. Franosch, 
Soft Matter \textbf{7}, 2076 (2011). 

\bibitem{FurstBook}
E. M. Furst and T. M. Squiers, 
\textit{Microrheology} 
(Oxford University Press, Oxford, 2017).

\bibitem{Granek11}
R. Granek,
Soft Matter \textbf{7}, 5281 (2011).

\end{thebibliography}
\end{document}